# Plasmonic metamaterial cloaking at optical frequencies

**F. Bilotti**[*], **S. Tricarico, and L. Vegni**

*Department of Applied Electronics, University "Roma Tre"*
*Via della Vasca Navale 84, Rome 00146, ITALY*
[*]*Corresponding author: bilotti@uniroma3.it*

**Abstract:** In this paper, we present the design of cylindrical and spherical electromagnetic cloaks working at visible frequencies. The cloak design is based on the employment of layered structures consisting of alternating plasmonic and non-plasmonic materials and exhibiting the collective behavior of an effective epsilon-near-zero material at optical frequencies. The design of a cylindrical cloak to hide cylindrical objects is firstly presented. Two alternative layouts are proposed and both magnetic and non-magnetic objects are considered. Then, the design of spherical cloaks is also presented. The full-wave simulations presented throughout the paper confirm the validity of the proposed setup, and show how this technique can be used to reduce the observability of cylindrical and spherical objects. The effect of the losses is also considered.

**OCIS codes:** 230.3205, 160.3918

## 1. Introduction

The growing interest in synthesizing a cloak of invisibility working in a desired frequency range has determined progressive efforts in order to obtain scattering-free and low-observability behaviors. A few different procedures to design electromagnetic cloaking devices have been proposed in the last few years by various groups worldwide [1-7]. Even if a real comparison between these different techniques in terms of performances is not so far possible according to the results presented in the technical literature, some significant advantages and drawbacks of each of these techniques can be pointed out. The approach presented in [4,5], essentially based on the use of homogeneous covers surrounding the object and characterized by atypical values (negative or close to zero) of their constitutive parameters, has undoubted advantages in terms of effective bandwidth and straightforward design. This is an important point, especially if we consider that, for instance, the technique proposed in [1-3], is based, instead, on the employment of inhomogeneous material cloaks with a certain profile of the constitutive parameters, which require a more complicated design.

Even if the approach presented in [4,5] seems to be promising, due to the employment of homogeneous covers, the design of actual materials exhibiting the required values of permittivity or permeability is, anyway, not trivial. In the microwave frequency range, an actual implementation of a cylindrical cloak has been recently proposed in [5]. In that case, the material with a negative real permittivity has been synthesized through a proper generalization of the parallel plate medium. If we would like to extend the cloaking approach proposed in [4,5] to the visible regime, we have at first to overcome some limitations. On one hand, in fact, the parallel plate medium cannot be used anymore at these frequencies, and, on the other hand, there is a lack of useful natural materials exhibiting the required permittivity function. As previously anticipated, in fact, for materials characterized by a close-to-zero real permittivity in the visible, the plasma frequency should rest in the visible, as well. Unfortunately, noble metals, such as gold and silver, have plasma frequencies at smaller wavelengths in the ultra-violet regime and, thus, exhibit a strong plasmonic (i.e. the real permittivity is strongly negative) behavior in the visible. For this reason, they are not useful for such purposes as such.

In this paper, we propose a possible way to apply the approach presented in [4,5] at optical frequencies, through the employment of proper layered structures made of plasmonic and non-plasmonic stacked materials.

## 2. Cylindrical cloaking devices: formulation and design

As previously anticipated, when we want to apply the procedure described in [4,5] to the design of actual cloaks working at optical frequencies, the main difficulty is represented by the synthesis of an artificial material with a plasma frequency laying in the visible, in order to get at those frequencies either an epsilon-near-zero (ENZ) behavior or a moderately negative value of the real part of the relative permittivity.

In order to overcome such a limitation, we may use the layered medium, which has been presented for a different application and in the case of a planar configuration in [8]. By stacking two different material slabs, if the thicknesses of the slabs are electrically small, it is shown in [8] that the resulting composite material is described through constitutive parameters which depend only on the ratio between the thicknesses of the two slabs and the constitutive parameters of the two different materials.

Now, let's suppose we want to cloak a cylindrical object with given permittivity and permeability ($\varepsilon_{obj}, \mu_{obj}$). Extending the formulation proposed in [8] to the case of the cylindrical geometry, we may have the two configurations depicted in Fig. 1. When moving from the planar geometry to the cylindrical one, in fact, it is possible to stack the two materials in two different ways. In both cases, the resulting composite material is anisotropic and the corresponding expressions of the entries of the permittivity tensor are given in the insets of the figure.

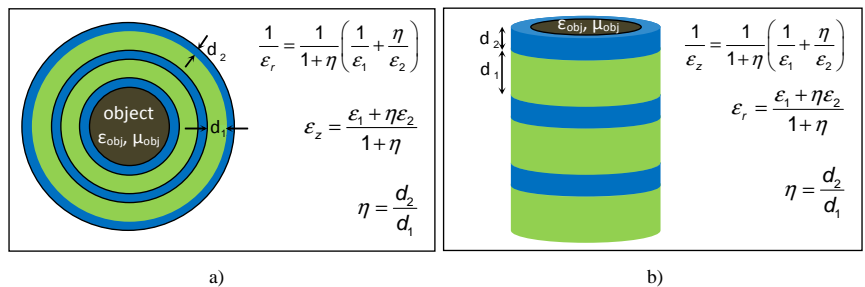

Fig. 1. a) Cylindrical cloak made of concentric shells of different materials. b) Cylindrical cloak made of stacked shells of different materials.

Depending on the polarization of the impinging wave and on the frequency range of interest, the designer can use either of the two configurations depicted in Fig. 1. Particularly, in order to get the desired value of the relative permittivity, the designer can count on a few degrees of freedom: the geometric parameters (i.e. the thicknesses of either the concentric shells of the setup in Fig. 1a or the slices of the setup in Fig. 1b) and the dispersive parameter behaviors of the two materials.

If the impinging wave is TM polarized (i.e. the electric field is along the axis of the cylinder), looking at the expression of the longitudinal effective permittivity in the insets of Fig. 1, it is clear that the configuration obtained by vertically stacking different material slices does not permit an easier implementation of an effective ENZ material in the visible. This configuration, in fact, has an effective relative permittivity dominated by the lower permittivity value, requiring that one of the two materials must have a relative permittivity less than one.

On the other hand, using a cylindrical cover made of concentric shells, the axial component of the permittivity tensor can be easily designed in order to have an ENZ behavior in the visible. If the two materials are a plasmonic material like silver (Ag) and a non-plasmonic material like silica ($SiO_2$), in fact, it is possible to obtain a close-to-zero real permittivity along the axis of the cylinder at optical frequencies simply by choosing the proper value for the ratio $\eta$ between the thicknesses of the shells made of the two different materials. Exploiting this configuration, thus, it is now possible to apply the formulation presented in [4,5] to reduce the total scattering cross-section (SCS) of a covered cylindrical object for the case of the TM polarization.

Let's consider a dielectric cylinder of radius $a$ = 50 nm and length $L$ = 500 nm with constitutive parameters $\varepsilon_{obj} = 2\varepsilon_0$, $\mu_{obj} = \mu_0$ illuminated by a plane wave with the electric field directed along the axis of the cylinder. Following the theoretical procedure presented in [4,5], it can be numerically found that, in order to reduce the scattering cross-section of the cylinder at the design frequency $f$ = 600 THz, an ideal homogeneous non-magnetic ($\mu_c = \mu_0$) metamaterial cover shell with radius $b = 1.8\ a$ (Fig. 2a) should have a permittivity $\varepsilon_c = 0.32\varepsilon_0$. In this case, the ideal reduction of the SCS of the covered cylinder in comparison to the case of the bare cylinder is *14 dB*.

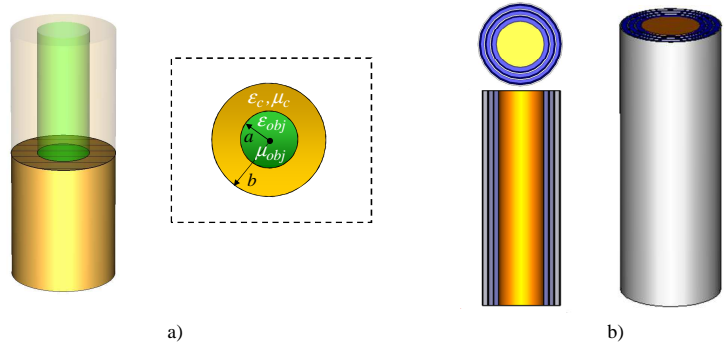

Fig. 2. a) Cylinder surrounded by an homogeneous cover. b) Cylinder surrounded by a cloak made of concentric shells of different materials.

Applying, now, the design formulas presented in the inset of Fig. 1 for the case of $SiO_2$-Ag concentric layers, the desired value of the effective permittivity ($\varepsilon_c = 0.32\varepsilon_0$) at the frequency $f$ = 600 THz is obtained with a thickness ratio $\eta = 0.21$. Given that the thickness of the cloak is limited by the radius value $b$ and the ratio $d_2/d_1 = \eta$ is now fixed, the choice of $d_1$ can be made in such a wise way to fill the cover with a sufficient number of concentric shells, according to the homogenization requirements. A layer of $SiO_2$ with $d_1$ = 10 nm is enough to obtain the desired dispersion near the working frequency (Fig. 3a), allowing the expected SCS reduction (Fig. 3b). The numerical results presented in Fig. 3b and all the other results presented in the following are obtained through full-wave numerical simulations performed with a numerical code based on the finite integration technique [9].

The reduction of the object observability around the desired frequency is quite evident. As expected, before the plasma frequency the observability of the object is partially increased, due to the strong plasmonic behavior of the Ag, while at higher frequencies the SCS of the covered object approaches the one of the bare cylinder.

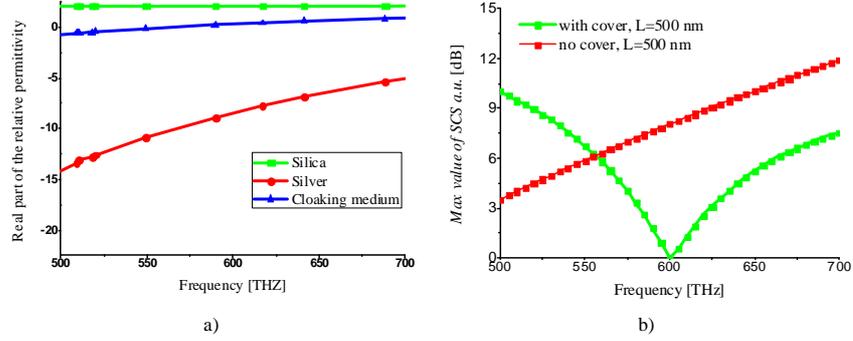

Fig. 3. a) Real part of the permittivity of Ag, SiO$_2$ and of the cloak effective medium (z-component). The cloak is designed for operation at $f$ = 600 THz and the thicknesses of the two material (Ag and SiO$_2$) shells are 2 nm and 10 nm, respectively. b) Maximum SCS of a cylindrical object ($L$ = 500 nm, $a$ = 50 nm, $\varepsilon_{obj} = 2\varepsilon_0$, $\mu_{obj} = \mu_0$) with and without the cloak.

In the frequency range of interest, Ag has relatively high losses. This explains the fact that the reduction of the SCS obtained in Fig. 3b is only *9 dB* instead of the *14 dB* expected from the theoretical calculations involving an ideal, lossless, unbounded and homogeneous cover. If we plot the ratio between the SCS of the bare cylinder and the one of the cylinder covered with an ideal cover as a function of the imaginary part of the cover relative permittivity $\varepsilon_c''$, we obtain the graph reported in Fig. 4. The reduction of the SCS in presence of the actual losses of the real-life layered material can be easily estimated from this graph and are in line with the results presented in Fig. 3.

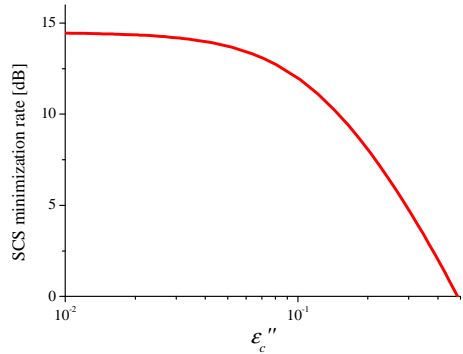

Fig. 4. Variation of the ratio between the SCS of the bare cylinder and the one of the cylinder covered with an ideal, lossy, homogeneous material, as a function of the imaginary part of the relative permittivity of the cover material. The real part of the permittivity is kept at the design value and the frequency is $f$ = 600 THz.

The pattern of the SCS in linear units reported in Fig. 5, clearly shows that the dominant scattering term for the bare cylinder is basically the dipolar one, due to its electrically small dimensions. Applying the layered cover, instead, the dipolar term is almost suppressed and the main contribution is given by an higher-order scattering mode.

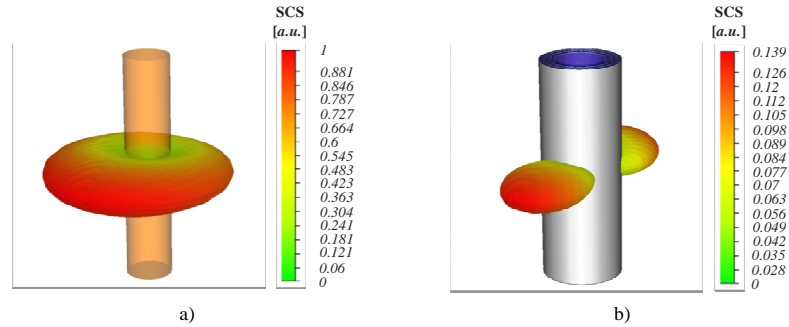

Fig. 5. a) SCS of the bare cylinder of Fig. 3 in linear units at $f = 600\ THz$. b) SCS of the cloaked cylinder of Fig. 3 in linear units at $f = 600\ THz$.

Looking at the bi-dimensional plots of the electric and magnetic field amplitude distributions at the cloak frequency (Figs. 6-7), the shadow effect is well evident in the case of the bare cylinder, while in the covered case it is visibly reduced.

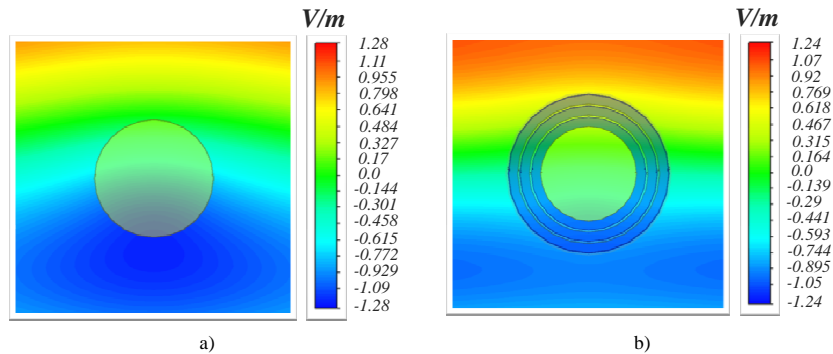

Fig. 6. Electric field amplitude at the cloak frequency $f = 600$ THz in the case of the bare cylinder (a) and the cylinder with the cloak (b).

The shadow reduction can be clearly appreciated in Fig. 8, where the isolines of the magnetic field at the cloak frequency for the bare cylinder and the cloaked one are shown. While in the first case the field lines are deformed by the presence of the obstacle, in the second one they are more similar to the field lines of a plane wave travelling unaltered through the object.

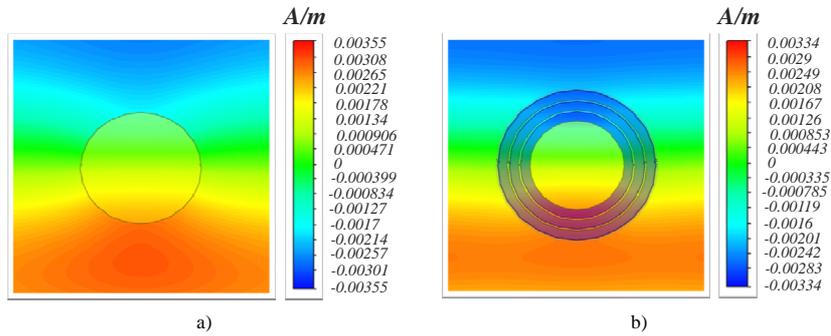

Fig. 7. Magnetic field amplitude at the cloak frequency $f$ = 600 THz in the case of the bare cylinder (a) and the cylinder with the cloak (b).

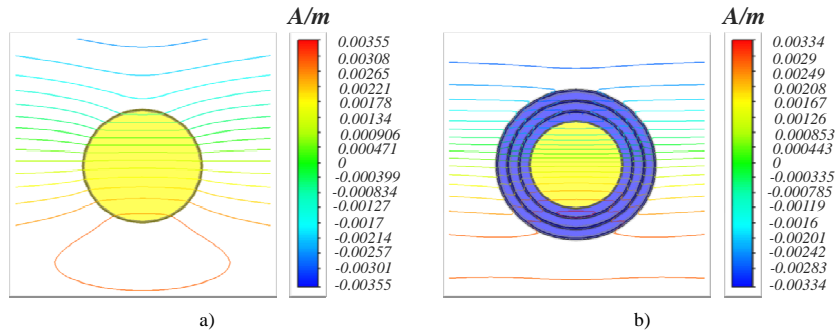

Fig. 8. Isolines of the magnetic field amplitude at the cloak frequency $f$ = 600 THz in the case of the bare cylinder (a) and the cylinder with the cloak (b).

The same approach can be used to cloak also an object with magnetic properties. Even if in the visible regime there are no natural materials exhibiting a magnetic response, it is, indeed, useful from a theoretical point of view to extend the same idea also to such materials. Let's consider now a cylinder of radius $a$ = 50 nm and length $L$ = 800 nm with constitutive parameters $\varepsilon_{obj} = 2\varepsilon_0$, $\mu_{obj} = 2\mu_0$ illuminated by a TM-polarized plane wave with the electric field directed along the cylinder axis. Again, using the numerical procedure in [4,5], it is easy to find that an ideal homogeneous non-magnetic ($\mu_c = \mu_0$) cover shell with the same radius $b$ = 1.8 $a$ and permittivity $\varepsilon_c = 0.1\varepsilon_0$ should be able to sensibly lower the SCS of the structure. In this configuration, using the same layer of $SiO_2$ with $d_1$ = 10 nm and Ag as the plasmonic material, the optimal value for the thickness ratio is $\eta$ = 0.21. In Fig. 9 it is shown the maximum of the SCS as a function of the frequency for the proposed cylinder with and without the cover. Also in this case, the reduction of the object SCS near the design frequency is quite evident.

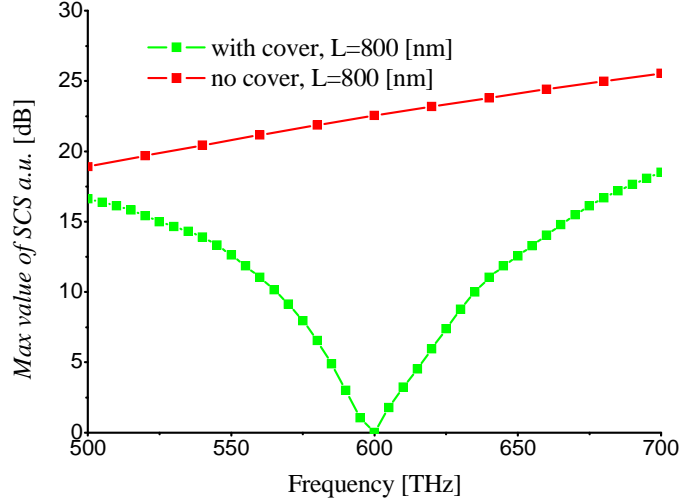

Fig. 9. Maximum of the SCS of a cylindrical object ($L = 800$ nm, $a = 50$ nm, $\varepsilon_{obj} = 2\varepsilon_0$, $\mu_{obj} = 2\mu_0$) with and without the cloak.

## 3. Spherical geometry: formulation and design

In this Section, we extend the previous results to the case of the spherical geometry. Due to the inherent anisotropic behavior of the layered medium (e.g. we cannot have the same values for the diagonal elements of the permittivity tensor), it is impossible to have a cloaking device working for both TM and TE polarizations at the same frequency. In the case of spherical layered structures, we could consider two different ways to stack the two materials. One possibility is to use concentric spherical shells and the other one is to use a stack of annular circular slices following the profile of the sphere. Anyway, the first setup is only suited for an impinging wave having a radial electric field. For this reason, in this paper, we propose to employ the second setup, which exhibits a given effective permittivity along the electric field direction of a TM-polarized impinging wave.

Accordingly, in the design reported in Fig. 10b, we have stacked the different material slices along the direction orthogonal to the incident electric field. The operation of such a configuration is substantially independent from the electric field direction in the plane of the layers, while we expect that the cloak should not work for the orthogonal polarization.

Let's consider, now, a silver spherical particle with radius $a = 50$ nm. Again, we use the homogeneous cover approach (Fig. 10a) in order to synthesize a spherical shell with a proper permittivity value. For an homogeneous non-magnetic ($\mu_c = \mu_0$) spherical shell with radius $b = 1.8\,a$ working at the frequency $f = 600$ THz, the required permittivity is $\varepsilon_c = 0.5\varepsilon_0$. Applying the same design formula as in the inset of Fig. 1a for the longitudinal permittivity and using again alternating layers of $SiO_2$ with thickness $d_1 = 10$ nm and Ag, the optimal value for the thickness ratio is found to be $\eta = 0.2$. In Fig. 11 it is shown the maximum of the SCS of the silver spherical shell with and without the cloak. Again, the observability of the object has been reduced at the design frequency by putting the cloak on it.

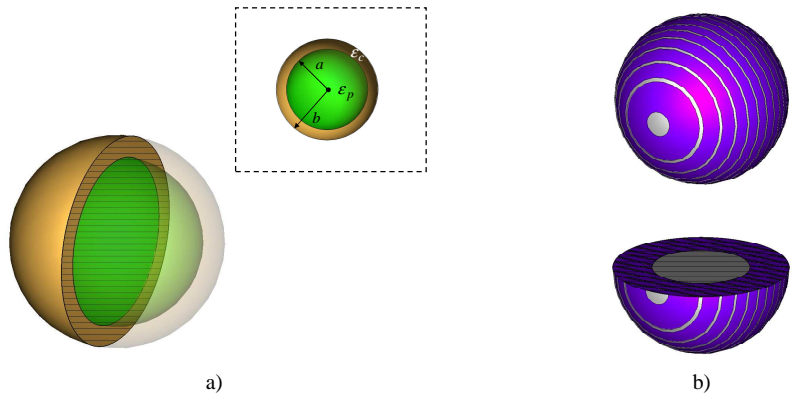

Fig. 10. a) A spherical particle surrounded by an homogeneous cover. b) A spherical particle covered with a cloak made of stacked layers of different materials.

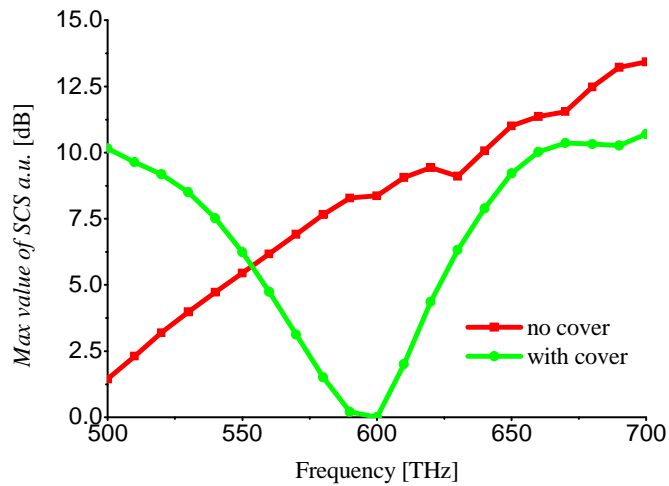

Fig. 11. Maximum of the SCS of a silver spherical particle ($a$ = 50 nm) with and without the cloak as a function of frequency.

The sensitivity of the cloaking device for variations of the geometrical parameters is shown in Fig. 12. As expected, a reduction of the ratio between the layer thicknesses corresponds to a lower-shift of the cloak operation frequency. Since the permittivity of $SiO_2$ is almost constant with the frequency, in fact, when lowering η, a higher (negative) value of the permittivity is required. The needed value can be found in the Ag permittivity dispersion at lower frequencies.

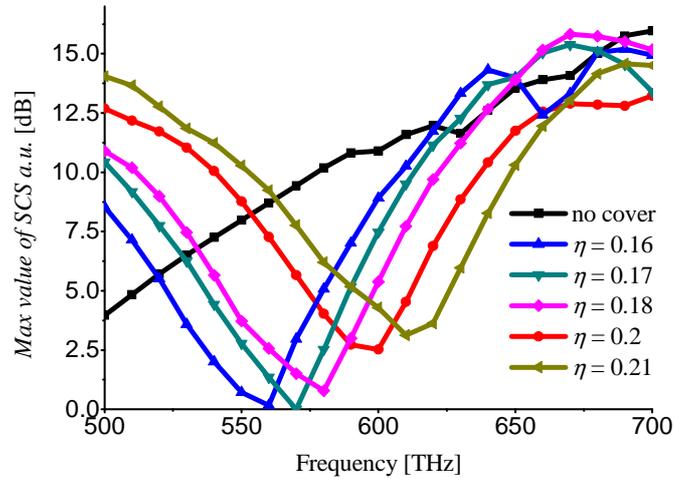

Fig. 12. Maximum of the SCS of a silver spherical particle ($a = 50$ nm) with and without the cloak as a function of frequency for different values of the ratio η.

As previously anticipated, the proposed setup is clearly independent from the orientation of the electric field in the plane of the layers, while it is expected not to work when the electric field is orthogonally directed with respect to the layers. In Fig. 13, we present the SCS of the same spherical particle used in the previous examples for differently polarized incident plane waves. When the electric field is orthogonal to the layers (Fig. 13b), the cloaking effect disappears, because the effective permittivity seen by the impinging wave (i.e. the orthogonal component of the permittivity) does not match the design value. In addition, since this permittivity is greater than $\varepsilon_0$, the total SCS is even bigger than the one of the bare silver sphere.

The SCS pattern obtained in the case of an impinging wave with the electric field parallel to the layers at the design frequency is reported in Fig. 14. The SCS reduction is rather evident, as can be easily remarked by the scale ranges.

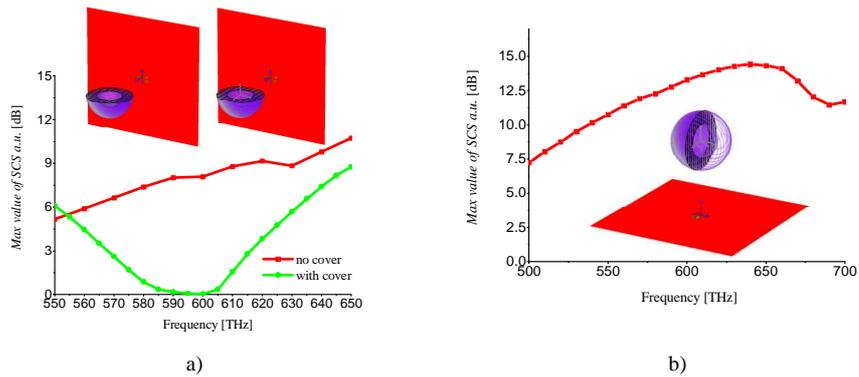

Fig. 13. a) Maximum of the SCS of a spherical silver particle ($a = 50$ nm) with and without the cloak for an impinging plane wave with the electric field parallel to the layers. b) Maximum of the SCS of the same particle with the same cloak for an impinging plane wave with the electric field orthogonal to the layers.

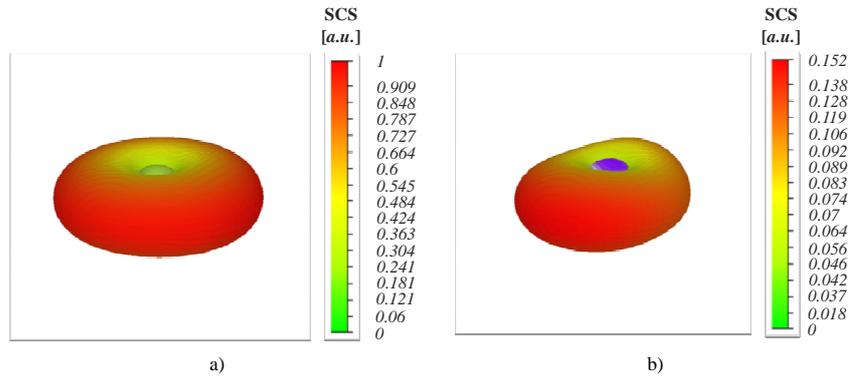

Fig. 14. a) SCS in linear units of the bare sphere at the cloak frequency $f = 600$ THz. b) SCS in linear units of the cloaked sphere at the same frequency. In both cases the impinging wave is polarized such that the electric field is parallel to the layers.

The bi-dimensional plots of the electric and magnetic field amplitude distributions at the design cloak frequency are reported in Figs. 15-16. From these graphs, it is well evident that in the case of the covered sphere the fields are noticeably less perturbed.

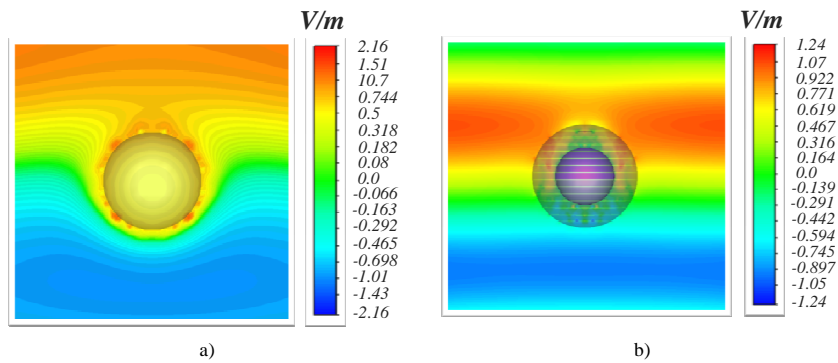

Fig. 15. a) Electric field amplitude at the cloak frequency $f = 600$ THz in the case of a bare sphere. b) Electric field amplitude at the cloak frequency $f = 600$ THz in the case of a sphere with the cloak.

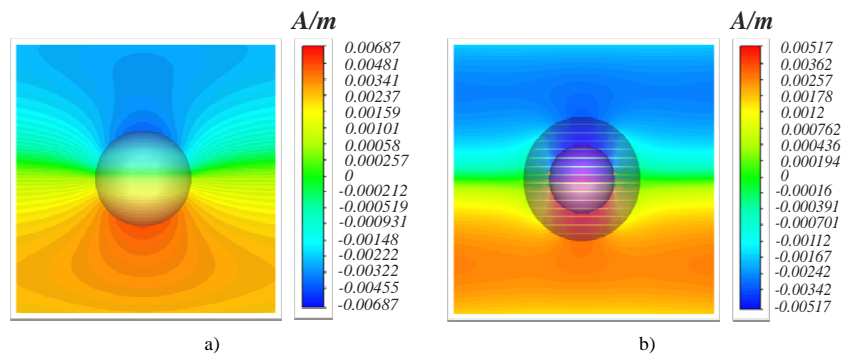

Fig. 16. a) Magnetic field amplitude at the cloak frequency $f = 600$ THz in the case of a bare sphere. b) Magnetic field amplitude at the cloak frequency $f = 600$ THz in the case of a sphere with the cloak.

The reduction of the shadow effect is clearly observed in Fig. 17, showing the isolines of the magnetic field at the cloak frequency for both the bare sphere and the cloaked one.

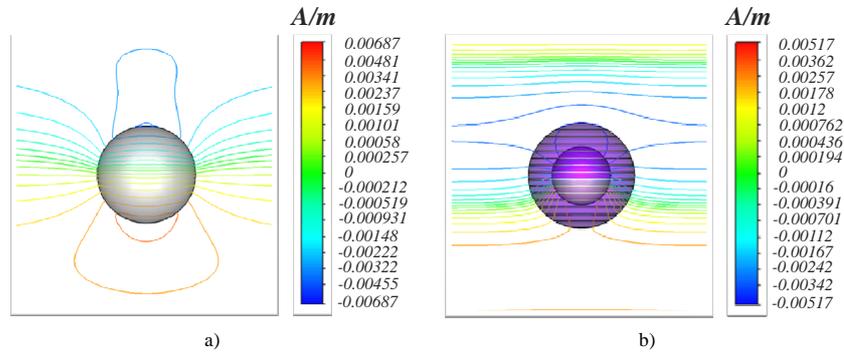

Fig. 17. a) Isolines of the magnetic field amplitude at the cloak frequency $f = 600$ THz in the case of a bare silver sphere. b) Isolines of the magnetic field amplitude at the cloak frequency $f = 600$ THz in the case of a silver sphere with the cloak.

## 4. Conclusion

In this paper, we have presented the design of electromagnetic cylindrical and spherical cloaks working at optical frequencies using layered structures of plasmonic and non-plasmonic materials. It is shown that a proper design of the effective layered medium, made of alternating layers of Ag and $SiO_2$, allows to obtain the desired ENZ behavior at the visible frequencies, allowing the actual realization of the cloaking devices. A few design examples of cylindrical and spherical cloaks have been proposed, considering also the effect of the losses and of possible magnetic objects. All the designs presented in the paper have been supported by full-wave simulations.

## 5. Acknowledgments

The authors would like to acknowledge the financial support of the following sources: ESA Ariadna Program, PRIN 2006, FP-7 ECONAM.